\documentclass[pre,amsmath,amssymb,floatfix]{revtex4}

\usepackage{graphicx}
\usepackage{latexsym}
\usepackage{amsfonts}
\usepackage{amssymb}
\usepackage{amsmath}

\newcommand{\bea}{\begin{eqnarray}}
\newcommand{\eea}{\end{eqnarray}}
\newcommand{\bef}{\begin{figure}}
\newcommand{\enf}{\end{figure}}
\newcommand{\ball}{\begin{array}{ll}}
\newcommand{\bacl}{\begin{array}{cl}}
\newcommand{\bal}{\begin{array}{l}}
\newcommand{\bac}{\begin{array}{c}}
\newcommand{\ea}{\end{array}}

\newcommand{\zbo}{{\mathbf{z}}}

\begin{document}

\title{Coarsening dynamics in a two-species zero-range process}

\author{S. Gro\ss kinsky$^1$ and T. Hanney$^2$\\
$^1$ Zentrum Mathematik, Technische
    Universit\"at M\"unchen, 85747 Garching bei M\"unchen, Germany \\
$^2$ School of Physics, University of Edinburgh,
  Mayfield Road, Edinburgh, EH9 3JZ, United Kingdom} 

\begin{abstract}
We consider a zero-range process with two species of interacting
particles. The steady state phase diagram of this model shows a
variety of condensate phases in which a single site
contains a finite fraction of all the particles in the
system. Starting from a homogeneous initial distribution, we study the
coarsening dynamics in each of these condensate phases, which is
expected to follow a scaling law. Random walk arguments are used to
predict the coarsening exponents in each condensate phase. They
are shown to depend on the form of the hop rates and on the
symmetry of the hopping dynamics.
The analytic predictions are found to be in good agreement with
the results of Monte Carlo simulations.
\end{abstract}

\maketitle

\section{Introduction}

Since the first observation of a condensation transition in the
homogeneous zero-range process (ZRP) \cite{E00} there has been a lot of
activity to further study this phenomenon on the level of the steady
state \cite{GSS}, and on the level of the relaxation dynamics
\cite{GSS,G03}. When the density of particles exceeds a critical
value, the system has been shown to phase separate into a homogeneous
background and a condensate which contains a finite fraction of all
the particles in the system. In the steady state the condensate
occupies only a single lattice site, and starting with homogeneous
initial conditions, the relaxation dynamics exhibit an interesting
coarsening phenomenon.

Besides being of interest in its own right as an example of a
condensation transition in an exactly solvable model, the phenomenon
is relevant in a more general context, providing a criterion for phase
separation in driven diffusive systems \cite{KLMST,ELMM}. The basic
condensation mechanism is by now well understood on a static and
dynamic level, but generalizations continue to be a topic of current
interest, such as coarsening behaviour on scale-free networks
\cite{noh}, processes with defect sites \cite{angel} or applications
to bipartite graphs \cite{pulkkinen}. Of particular
interest are generalisations to two-species zero-range processes with
two conservation laws, which also exhibit condensation and have a much
richer stationary phase diagram than that of the single species system
\cite{EH03,HE04}. Indeed, while the stationary and dynamical
properties of one-dimensional driven diffusive systems with one
species of particles are relatively well understood, much less is
known about the properties, and in particular the dynamical
properties, of driven systems with two or more species of conserved
particles (see \cite{S03} for a recent review).

This paper provides a first analysis of the
coarsening dynamics of a two-species zero-range process.
%providing a first analysis of the coarsening behaviour in a two species
%system.
We generalize the arguments in \cite{GSS} for a single species
system, which turns out to
be far from straightforward since several new effects have to be taken
into account, effects due to the coupled dynamics of the two particle
species. The model is chosen such that all the expected new features
can be observed while the steady state is exactly solvable. 
%So this
%case study can serve as a prototype for the analysis of other
%two-component systems. 
In addition to this theoretical interest, the results are relevant for physical
realisations of two species zero-range processes, which can
be found for example in shaken bidisperse granular systems
\cite{MMWL02} and models of directed networks \cite{DMS03}.

In Section II, we define the model which is a generalisation of the
model considered in \cite{HE04}, recap some known results for the
steady state and give the phase diagram. In Section III we state the
expected scaling behaviour for the coarsening regime and explain the
random walk arguments for its analysis. The main results of the paper
are derived in Section IV: scaling laws for the time evolution of the
mean condensate size for all regions of the phase diagram,
generalizing the derivation in \cite{GSS}.  The predictions are
compared to Monte Carlo simulation data and we find good agreement. We
conclude in Section V and include a discussion of finite size effects
in an appendix.

\section{Model}

\subsection{Definition and steady state}

We define the two-species zero-range process on a one-dimensional
lattice containing $L$ sites with periodic boundary conditions. On
this lattice, there are $N_1$ particles of species $1$ and $N_2$
particles of species $2$.
%The density of particles of species $i$ is $\rho_i = N_i/L$.
A site with occupation numbers $k_1$ and $k_2$ for
species 1 and 2 respectively, loses a particle of species 1 with rate
$g_1(k_1, k_2)$ and of species $2$ with rate $g_2(k_1, k_2)$.  For
simplicity we assume that particles hop to their nearest neighbour
site to the right, although our results also apply for more general
hopping of finite range.

The steady states for this model with general $g_1(k_1, k_2)$ and
$g_2(k_1, k_2)$ have been characterised in \cite{EH03,GS03} and we now
summarise the main points. We denote a particle configuration by ${\bf
  k}=k_{1,1} ,k_{2,1};\ldots;k_{1,L},k_{2,L}$. The steady state
probabilities assume a factorised form  
\bea \label{nu^L}
\nu_\zbo^L ({\bf k}) = \prod_{x=1}^{L} \nu_\zbo (k_{1,x} ,k_{2,x})\;,
\eea
provided the hop rates satisfy the constraint  
\begin{equation} \label{constraint}
\frac{g_1(k_1, k_2)}{g_1(k_1, k_2-1)} = \frac{g_2(k_1,
  k_2)}{g_2(k_1-1, k_2)}\;,
\end{equation}
for all $k_1 ,k_2 \geq 1$.
% they can be written as
%\begin{equation} \label{integrator}
%g_1 (k_1 ,k_2 )=\frac{f(k_1, k_2)}{f(k_1 -1, k_2 )}\quad ,\qquad g_2
%(k_1 ,k_2 )=\frac{f(k_1, k_2)}{f(k_1 , k_2 -1)}  
%\end{equation}
%with an integrating function $f(k_1 ,k_2 )$. For given rates it is
%uniquely determined up to a multiplicative constant and can be
%represented in several ways using (\ref{integrator}), one of which is
The single-site distribution has the form
\begin{equation}\label{stst}
\nu_\zbo (k_1 ,k_2)=\frac{1}{Z(\zbo )}\, f(k_1 ,k_2 )\, z_1^{k_1}
z_2^{k_2} \;,
\end{equation}
where $f(k_1 ,k_2 )$ is a stationary weight which can be written
\begin{equation} \label{representation}
f(k_1 ,k_2 )=\prod_{i=1}^{k_1}\frac{1}{g_1 (i,0)}\
\prod_{j=1}^{k_2}\frac{1}{g_2 (k_1 ,j)}\;.
\end{equation}
Here $\zbo =(z_1 ,z_2 )$, $z_i \geq 0$ play the role of fugacities for
each species, in that they are chosen to fix the particle densities
$\rho_i =\langle N_i \rangle_\nu /L$ for species $i=1,2$, i.e.\ the
expected number of particles per site in the steady state. Thus we are
working in a
grand canonical ensemble, which is normalised by the single site
partition function $Z(\zbo )$. This steady state can be directly obtained by
substitution into the balance condition for the steady state
probability that the system is in a configuration ${\bf k}$.
%It is elementary to check that $f(k_1 ,k_2 )$ serves as the stationary
%weight of a factorised steady state $\nu_\zbo^L$ with  
%It was
%found that for the rates (\ref{rats}), the steady state factorises
%into a product
%\bea \label{nu^L}
%\nu_\zbo^L ({\bf k}) = \prod_{x=1}^{L} \nu_\zbo (k_{1,x} ,k_{2,x})\;,
%\eea
%where we denote a particle configuration by ${\bf k}=k_{1,1}
%,k_{2,1};\ldots;k_{1,L},k_{2,L}$, 
%single-site distribution
%$\nu_\zbo (k_1 ,k_2 )$ has the form   
%\begin{equation}\label{stst}
%\nu_\zbo (k_1 ,k_2)=\frac{1}{Z(\zbo )}\, f(k_1 ,k_2 )\, z_1^{k_1}
%z_2^{k_2} \;.
%\end{equation}

We remark that one gets the same steady state if the hopping
dynamics are symmetric, rather than asymmetric as defined above. A
useful property of the steady state \cite{EH03,GS03} is that the
expectation value of the hop rate of species $i$, denoted by $\langle
g_i \rangle_\nu$, is equal to $z_i$. Thus $\langle g_i \rangle_\nu$ is
a translation invariant quantity; this is obvious in the case of
asymmetric dynamics where $\langle g_i \rangle_\nu$ is the current,
but less obvious in the case of symmetric dynamics with vanishing
current, where $\langle g_i \rangle_\nu >0$.

We are interested in the coarsening dynamics of the model in various
phases that arise for a particular choice of rates, namely 
\bea
\label{rats} g_1(k_1,k_2) &=& \left(
\frac{1+b/(k_1+1)^\gamma}{1+b/k_1^\gamma}
\right)^{k_2}(1+c/k_1)\;,\nonumber\\
g_2(k_1,k_2)&=&1+b/(k_1+1)^\gamma\;, 
\eea 
where $g_1 (0,k_2 )=g_2 (k_1 ,0)=0$ and $b,c,\gamma>0$. It is easy to
check by substitution that these rates satisfy the constraint
(\ref{constraint}). The stationary weights, obtained from
(\ref{representation}), are given by
\begin{equation}
f(k_1 ,k_2 )=\frac{k_1!}{(1+c)_{k_1}}\Big(1+\frac{b}{(k_1
  +1)^\gamma}\Big)^{-k_2} \ , 
\end{equation}
where $(a)_k = \prod_{i=0}^{k-1}(a+i)$ is the Pochhammer symbol.
The single-site partition function is given by  
\bea\label{partition}
Z(\zbo ) =\sum_{k_1 ,k_2 =0}^\infty f(k_1 ,k_2 )\, z_1^{k_1}
z_2^{k_2} =\sum_{k_1 =0}^\infty z_1^{k_1}
\frac{(k_1 +1)^\gamma +b}{(1-z_2 )(k_1 +1)^\gamma +b}\ \frac{k_1
!}{(1+c)_{k_1}}\ .  
\eea 

We make the choice (\ref{rats}) in order to study the behaviour when
the dynamics of one of the particle species, here species 2, depends
only on the number of particles of the other species at the departure
site. So condensation of species 2, when it occurs, is induced by the
presence of species 1 particles, which can be interpreted as an
evolving disordered background as discussed for a specific case in
Section IV.C. The $k_2$ dependence in $g_1$ is then determined by the
constraint (\ref{constraint}). The second factor $(1+c/k_1)$ in $g_1$
could be replaced by any function of $k_1$ and the steady state will
still factorise. The form we have chosen is the simplest form of the
hop rate for which the single-species zero-range process exhibits a
condensation transition for $c>2$ at a finite critical
density of particles \cite{E00}. Thus the parameter $c$ can be tuned to allow
also condensation of species 1 particles, which influences the phase
diagram of the process as discussed in Section II.B. We remark that
the existence of condensation transitions and any subsequent
coarsening behaviour depend only on the asymptotic forms of the rates.
Other choices of this second factor, with different asymptotic
properties, lead either to no transition of species 1 particles if it
is non-decreasing or
tends to a constant faster than $2/k_1$ as $k_1 \to \infty$, or to
condensation at any density (where the fraction of particles in the
condensate is equal to one) if it tends to zero as $k_1 \to \infty$.
Thus (\ref{rats}) are basic rates which capture two different mechanisms of
condensation transition (i.e.\ induced and autonomous) and the most
interesting coarsening behaviour that we expect to observe while the
steady state remains exactly solvable.

\subsection{Stationary phase diagram}

\begin{figure}
\begin{center}
\includegraphics[width=0.33\textwidth]{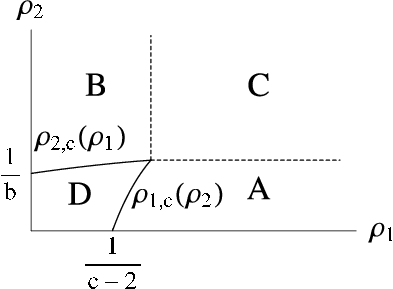}\hfill\includegraphics[width=0.33\textwidth]{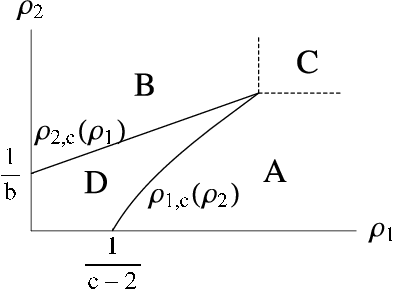}\hfill\includegraphics[width=0.33\textwidth]{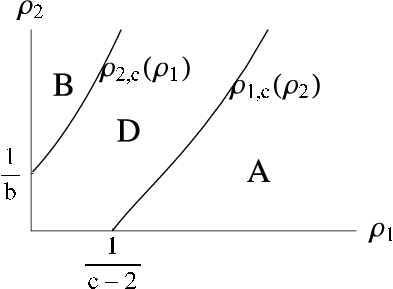}\\
$\gamma =0.5$\hspace*{0.25\textwidth} $\gamma
=1$\hspace*{0.25\textwidth} $\gamma =2$\\ 
\caption{Phase diagram for the choice of rates (\ref{rats}) with
$b=1$, $c=4$ and several values of $\gamma$. For $\gamma =0.5$ and
$\gamma =1$ it is $c>\max (2+\gamma,1+2\gamma)$ and region C exists.
Whereas for $\gamma =2$ we have $c<\max (2+\gamma,1+2\gamma)$ and
region C does not exist. See text for details.} 
\label{PD}
\end{center}
\end{figure}

The range of possible fugacities is given by the domain of convergence
of the partition function $Z(\zbo )$ given in (\ref{partition}). In
the present case the maximal fugacities are $z_1 =1$ and $z_2 =1$ and
when one or both of the fugacities are maximal we use the notation
$\zbo =\zbo_c$. The phase diagram in terms of the particle densities
$\rho_1$ and $\rho_2$ has been derived in \cite{HE04,G}.
For the grand canonical ensemble (\ref{stst}) the densities are given by
\bea
\rho_i =z_i \;\frac{\partial \ln Z(\zbo )}{\partial z_i}\ ,\quad i=1,2\ ,
\eea
and thus the
convergence properties of the partition function at the maximal
fugacities determine whether or not the critical
densities $\rho_{i,c} :=\rho_i \big|_{z_i =1}$ are finite or infinite.
%$\rho_{1,c}$ and $\rho_{2,c}$, characterized by $z_1 =1$ and
%$z_2 =1$, respectively. 
In general, $\rho_{1,c}$ can depend on
$\rho_2$ i.e.\ $\rho_{1,c} = \rho_{1,c}(\rho_2)$ and vice versa. If
$\rho_i \leq \rho_{i,c}$ for $i=1,2$, both species are in a fluid
phase corresponding to a factorised steady state $\nu_{\zbo}$ as given
in (\ref{stst}). In the phase diagram shown in Figure \ref{PD} this
region is denoted by D. If the particle density $\rho_i$ of
either species $i=1,2$ exceeds its critical value $\rho_{i,c}$,
species $i$ condenses: the system phase separates into a homogeneous
background fluid phase with distribution $\nu_{\zbo_c}$, and a
condensate which contains the $(\rho_i-\rho_{i,c})L$ `excess'
particles of species $i$. In a typical stationary configuration this
condensate occupies a single, randomly located site.

Depending on the values of $c$ and $\gamma$ in (\ref{rats}) the
following phases appear in the phase diagram in addition to the fluid
phase D:
\begin{itemize}
\item In region A, the fugacities are $\zbo_c=(1,z_2)$ with $z_2<1$. The
species 1 particles condense and the species 2 particles form a
fluid. The particle densities in the background phase are
$(\rho_{1,c},\rho_2 )$.

\item In region B, the fugacities are $\zbo_c =(z_1 ,1)$ with $z_1
<1$, species 2 condenses and the background
particle densities are $(\rho_1 ,\rho_{2,c} )$. As an additional
point, the site containing the condensate of species 2 particles also
contains $\mathcal{O}(L^{1/(1+\gamma)})$ species 1 particles \cite{EH03}.

\item In region C, $\zbo_c=(1,1)$. A single site contains condensates of
both species and the
background densities are $(\rho_{1,c} ,\rho_{2,c} )$.
\end{itemize}

The phase diagram shown in Figure \ref{PD} is richest when $c>\max
(2+\gamma,1+2\gamma)$, where all three regions are found.
For $2<c\le\max (2+\gamma, 1+2\gamma)$ the phase diagram contains only
the phases A, B and D, and for $c\leq 2$ only phases B and D remain.

So far we have discussed straightforward generalisations of previously
known results. We now turn to the main aim of this work, which is to
study the coarsening dynamics of the two-species zero-range process
leading to each of the condensate phases A, B and C.

\section{Coarsening}
In the following we use the symbol $\approx$ to denote asymptotic
expansions in the thermodynamic limit $L\to\infty$ with fixed
particle densities, i.e.\ $N_1 =[\rho_1 L]$ and $N_2 =[\rho_2 L]$. 
If the terms in the expansion are only given up to
a constant factor we use the symbol $\sim$ instead.

\subsection{Relaxation dynamics}

In this section we outline the arguments used to
describe the coarsening dynamics in the condensate phases.
Starting from an initially uniform distribution of
particles, the dynamics of the condensation can be divided
into three regimes:
\begin{itemize}
\item[(i)] nucleation, during which excess particles of
either species accumulate at several randomly located sites, which we
call \textbf{cluster sites}. Each contains $\mathcal{O}(L)$
particles, so there are $\mathcal{O}(1)$ cluster sites, separated by a
typical distance of order $L$. At the remaining sites, which we call
\textbf{bulk sites}, the system relaxes to its steady state distribution
$\nu_{\zbo_c}$.
\item[(ii)] coarsening, during which the cluster sites exchange
  particles through the bulk. This leads to the growth of large
  condensates at the expense of smaller ones and a decrease in the
  number of cluster sites.

\item[(iii)] saturation, where eventually only two cluster sites
  remain due to the finite size of the system. In this regime
  the dynamics, under which the system reaches a typical steady state
  configuration with a single cluster site, is different from the
  coarsening dynamics (cf.~\cite{GSS}).
\end{itemize}
Physically, the most interesting is the coarsening regime.  Here,
large condensates gain particles at the expense of smaller ones which
causes some condensates to disappear. This in turn leads to a decrease
in the number of cluster sites and hence an increase of the mean
condensate size $m_i(t)$, defined as the number of particles of
species $i=1,2$ at cluster sites divided by the number of cluster
sites at time $t$.  In the limit $t\to\infty$, $m_i(t)$ converges to
its steady state value $(\rho_i -\rho_{i,c})L$, the number of excess
particles of species $i$ in the system. Within the coarsening regime
the increase of the mean condensate size is expected to
follow a scaling law, $\big\langle m_i(t) \big\rangle_L \sim
t^{\beta_i}$. Moreover, on a certain time scale $\tau = \tau (L)$,
the growth of the normalised mean condensate size is expected to
be independent of the system size $L$,
\begin{equation}\label{sclaw}
\frac{\big\langle m_i(t) \big\rangle_L }{(\rho_i -\rho_{i,c})L}\sim
(t/\tau )^{\beta_i}\;. 
\end{equation}
%with a certain time scale $\tau =\tau (L)$, which depends on the
%system size $L$ (and other system parameters). 
Therefore the scale $L$ of the mean condensate size and the time scale $\tau$
are connected via $L\sim \tau^{\beta_i}$, and thus $\tau\sim L^{1/\beta_i}$.
The angled brackets
$\langle ..\rangle_L$ denote an ensemble average in a finite system of
size $L$, starting with a homogeneous distribution of $N_j =[\rho_j
L]$ particles for both species $j=1,2$. This is in contrast to the
steady state expectation denoted by $\langle ..\rangle_\nu$.  The
scaling law (\ref{sclaw}) defines the exponent $\beta_i$, which may
depend on the particle species $i$. In general one could choose
different observables to monitor the coarsening process, such as the
square sum of occupation numbers. But in our case the mean condensate
size is a natural choice, since it is directly accessible by our
arguments given below.

We remark that the scaling law (\ref{sclaw}) is of the same form as
that which describes the growth of characteristic length
scales in phase ordering dynamics \cite{B94}.
More precisely one can define a scaling function
\bea\label{sclaw2}
h_i (t'):=\lim_{L\to\infty} \frac{\big\langle m_i(t'\tau ) \big\rangle_L
}{(\rho_i-\rho_{i,c})L}\;,
\eea
for all $t' \geq 0$. With the appropriate time scale $\tau$,
which will be derived in the next section for the various phases,
$h_i$ is expected to be a non-degenerate, smoothly increasing function
with the asymptotic properties
\bea
h_i (t')=\mathcal{O}\big( t'^{\beta_i }\big)\ \ \mbox{for }t'\to
0\qquad\mbox{and}\qquad\lim_{t'\to\infty} h_i (t')=1\ .
\eea
So for small $t'$ the coarsening regime is described by a power law
(\ref{sclaw}) which we study in the following. For $t'\to\infty$ the
system saturates and $h_i$ converges to its maximal value $1$. We do
not further discuss the behaviour in this regime, this has been done
for a single species system in \cite{GSS}.

\subsection{Random walk arguments}

In the following, our aim is to estimate the exponents $\beta_i$ in
each of the condensate phases A, B and C of the model defined by the
rates (\ref{rats}). This is achieved by adapting the random walk
arguments given for the coarsening dynamics of the one-species model
\cite{GSS}. They are based on two major assumptions, which are
self-consistent and confirmed by simulation data.
\begin{itemize}
\item[(A1)] Separation of time scales:\\
The nucleation process is very fast so that during the coarsening
regime the bulk sites have already relaxed to the steady state distribution
$\nu_{\zbo_c}$.
\end{itemize}
%In the nucleation regime we expect exponential relaxation on a time
%scale of order $L$.
Within the coarsening regime the system can
therefore be separated into a stationary bulk and a finite number
of isolated cluster sites. On top
of stationary hop rates $\langle g_i \rangle_\nu =z_{i,c}$, cluster
sites of species $i$ exchange particles through the bulk on a slower
time scale, given below. The bulk can be seen as a homogeneous medium
through which these excess particles perform a biased random walk, and
the cluster sites as boundaries where they enter and exit.
\begin{itemize}
\item[(A2)] Independence of excess particles in the bulk:\\
The excess particles exchanged by cluster sites perform independent (biased)
random walks through the bulk on their way to the next cluster site and do not
effect the bulk distribution $\nu_{\zbo_c}$.
\end{itemize}
%Based on these assumptions the dynamics within the coarsening regime
%is described below. 
This is justified below by noting that the average density of excess
particles in the bulk vanishes for $L\to\infty$. 

The random walk argument then proceeds as follows. We consider the case where
one species $i=1$ or $2$ condenses. The rates we consider decay as
$g_i -1 \sim k_i^{-\alpha}$ with $0<\alpha\leq 1$ (see Section IV) and the
average hop rate in the bulk is $z_{i,c} =1$. So the effective rate at
which cluster sites with $k_i \sim L$ lose particles to the bulk is
$g_i -z_{i,c}\sim L^{-\alpha}$.
These excess particles perform a biased random walk through the
bulk with drift
\bea\label{drift}
\langle g_i \,|\, k_i >0\rangle_\nu -\langle g_i \rangle_\nu
=\frac{\langle g_i \rangle_\nu}{1-\nu_{\zbo_c} (k_i =0)}-\langle g_i
\rangle_\nu =\frac{1}{1-\nu_{\zbo_c} (k_i =0)}-1\ .
\eea
Since $\nu_{\zbo_c} (k_i =0)>0$ this is positive and independent of $L$.
Thus the time it takes an excess particle to reach a neighbouring cluster
site scales as the typical distance between cluster sites which is
$\mathcal{O}(L)$. So $n$ independent excess particles exit the bulk
with rate $\mathcal{O}(n/L)$ which has to balance the entry rate of
order $L^{-\alpha}$. Hence, the number of excess particles in the bulk
scales as $\mathcal{O}(L^{1-\alpha})$ which grows only sublinearly
with $L$ for $0<\alpha\leq 1$, consistent with (A2).

In this balanced situation the time scale on which cluster sites
exchange single particles through the bulk is $\mathcal{O} (L^\alpha
)$. The time scale on which cluster
sites exchange a finite fraction $\sim L$ of their particles is thus
$\mathcal{O} (L^{1+\alpha})$. Since by definition the number of
cluster sites during coarsening is of order $1$, this sets the
coarsening time scale $\tau$ and the coarsening exponent $\beta_i$ in
(\ref{sclaw}) to be
%From the above statements, we also conclude that the motion of excess
%particles through the bulk does not limit the coarsening time scale
%since the typical time it takes particles to escape from cluster sites
%is of the same order in $L$ as the typical time excess particles spend
%in the bulk; the coarsening time scale is only determined by the
%effective rates at which cluster sites lose particles. 
%The coarsening is driven by a decrease of cluster sites and 
%The time scale for a cluster to lose all of its particles is 
\bea \label{timescale}
\tau\sim L^{1+\alpha}\ ,\quad\beta_i =\frac{1}{1+\alpha}\ .
\eea
%Since by definition the number of cluster sites during coarsening is
%of order $1$ this sets the time scale for the coarsening regime to be
%$\tau \sim L^{1+\alpha}$, consistent with (A1). In turn, solving
%(\ref{timescale}) for $k_i (t)$ and normalizing by 
%$(\rho_i -\rho_{i,c})L$, we recover the expected scaling law
With the above considerations we can give an additional motivation for
the scaling behaviour (\ref{sclaw}). We have seen that the
rate at which cluster sites exchange particles through the bulk
depends on their size as $k_i^{-\alpha}$. Thus, in a very rough
approximation, the time derivative of
the average condensate size $\big\langle m_i(t) \big\rangle_L$ should
be proportional to the average exchange rate of excess particles,
\bea
\frac{d\big\langle m_i(t) \big\rangle_L}{dt}\sim \big\langle m_i(t)
\big\rangle_L^{-\alpha}\ .
\eea
As the solution we recover the scaling law
(\ref{sclaw}) with exponent $\beta_i$ as given above.
This is of course not a strict argument and should not be understood as a
derivation of the scaling law.

Compared to the bulk dynamics the coarsening is a very slow process
and typical configurations with cluster sites on top of a stationary
background are quasi-stationary, i.e.\ within times of order $1$ the
configurations at cluster sites do not change on average. To leading
order in $L$ the dynamics on cluster sites have to be compatible with
the stationary bulk dynamics. By compatibility we mean that the
translation invariance of $\langle g_i \rangle_{\nu}$ implies that
%$\langle g_i \rangle_{\nu}$
it must be the same for all sites (both
cluster sites and bulk sites) in the system. For two-component
systems, this induces consistency
relations between the occupation numbers $k_1$ and $k_2$ on cluster
sites, a fact that will often be used below. If these relations are
not fulfilled, the configuration is not quasi-stationary in the above
sense and changes on time scales of order $1$ through interaction with
the bulk.

\section{Coarsening scaling laws}

\subsection{Theoretical predictions}

We now apply the arguments described above to the model at hand and
explain additional effects due to the presence of two species of
particles.

\subsubsection{Phase A}

In phase A, $\rho_1 >\rho_{1,c}$ and only the first species
condenses. There are $(\rho_1-\rho_{1,c})L$ excess particles of
species 1
in the system and at the cluster sites $k_1 =
\mathcal{O}\big( (\rho_1-\rho_{1,c})L\big)$, while $k_2$ remains finite in the limit
$L\to \infty$, which is justified below by compatibility with the
bulk. Hence, at the cluster sites the rates (\ref{rats}), up to first
order in $k_1$, are given by
\begin{equation}\label{ratesa}
g_1(k_1, k_2) \approx 1+c/k_1\;,\quad g_2(k_1, k_2) \approx 1+b/k_1^\gamma\;.
\end{equation}
Thus the coarsening of the species 1 particles is independent of the
second species: the net rate at which particles leave a cluster site
is $g_1 -1=c/k_1$.
Following the arguments leading to (\ref{timescale}) in Section III.B
the coarsening time scale is thus
\begin{equation}
\tau_A \sim [(\rho_1-\rho_{1,c})L]^2 \;,
\end{equation}
and we expect that the normalised mean condensate size grows like
\begin{equation}\label{scalea}
\frac{\langle m_1(t) \rangle_L }{(\rho_1-\rho_{1,c})L} \sim (t/\tau_A
)^{1/2}\; \quad {\rm i.e.} \quad \beta_1 = 1/2 \;.
\end{equation}
This recovers the known coarsening of the condensate in the
one-species ZRP where particles hop with rate $1+c/k$ \cite{GSS,G03}.

Further, because the jump rates of both species are coupled, the
presence of a species $1$ condensate influences the distribution $P$
of the species $2$ particles on the cluster site: Since cluster sites
and bulk have to be compatible, $g_2$ on the cluster site has to be
equal to the bulk steady state current $\langle g_2 \rangle_\nu =z_2 <1$, and
using (\ref{ratesa}) we have
\begin{equation}
\langle g_2 \rangle_\nu \approx (1+b/k_1^\gamma )P(k_2>0)\ \approx\ P(k_2 >0)\;.
\end{equation}
Therefore $P(k_2 =0)\approx 1-\langle g_2 \rangle_\nu$. This is non-zero, but
smaller than the expected bulk value, which contains an extra positive
contribution due to $b/k_1^\gamma =\mathcal{O}(1)$.

\subsubsection{Phase B}

In phase B, $\rho_2 >\rho_{2,c}$ and the second species condenses. The
number of particles at a cluster site is $k_2 =
\mathcal{O}\big( (\rho_2-\rho_{2,c})L\big)$. Now, using (\ref{rats}), the hop
rate $g_1$ of the first species at a cluster site vanishes in the
limit $L\to\infty$ if $k_1 =\mathcal{O}(1)$. But since in the bulk the
mean hop rate of the first species is given by its steady state value
$\langle g_1 \rangle_\nu =z_1 \in (0,1)$ for $\rho_1 >0$, $k_1$ has to
be large at cluster sites. Thus, considering $k_1$ large in
(\ref{rats}), the hop rate of species 1 particles at cluster sites
becomes
\begin{equation}\label{ratesb1}
g_1(k_1,k_2) \approx {\rm exp}(-b \gamma k_2/k_1^{1+\gamma}) (1+c/k_1)\;, 
\end{equation}
which should be consistent with
the expected bulk value $\langle g_1 \rangle_\nu
=z_1$. This compatibility requirement leads to
\bea\label{rel}
k_1^{1+\gamma} \approx -b \gamma k_2/{\rm ln} z_1 \; ,
\eea
and this relation between $k_1$ and $k_2$
is dynamically stable since on cluster sites 
\bea
\partial_{k_1} (g_1 (k_1 ,k_2 )-z_1 )\big|_{k_1^{1+\gamma} \approx -b
  \gamma k_2/{\rm ln} z_1} \approx \frac{-(1+\gamma )\, z_1 \,\log
  z_1}{k_1} >0\;.
\eea
So cluster sites where $k_1$ is too small gain species 1 particles
from the bulk (or if $k_1$ is too high they are lost to the bulk), and
thus any perturbation of the relationship (\ref{rel}) is 
driven towards this stable form on intermediate time scales. Hence
cluster sites at which (\ref{rel}) is satisfied dominate
the coarsening and the hop rate of the second species can be written 
\begin{equation}\label{ratesb2}
g_2(k_1,k_2) = 1+b/(k_1 +1)^\gamma \approx 1+\left(\frac{-b^{1/\gamma} \,{\rm
      ln}z_1}{\gamma k_2}\right)^{\frac{\gamma}{1+\gamma}}\;.
\end{equation}  
Therefore particles of species 2 escape from a cluster site at a net
rate proportional to $1/k_2^{\gamma/(1+\gamma)}$ and we can repeat the
arguments given in Section III.B to deduce the coarsening time scale 
%is
%determined by the time it takes a cluster site to lose all of its
%particles, $t(k_2) \sim k_2^{1+\gamma/(1+\gamma)}
%=k_2^{(1+2\gamma)/(1+\gamma)}$. Since we are only interested in the
%scaling with $k_2$ we omit the proportionality constant, which has a
%rather lengthy dependence on $\gamma$, $b$ and $z_1$. The time scale
%$\tau_B$ is then given by
\begin{equation} 
\tau_B \sim [(\rho_2-\rho_{2,c})L]^{\frac{1+2\gamma}{1+\gamma}} \;.
\end{equation}
The normalised mean condensate size grows like
\begin{equation}
\frac{\langle m_2(t) \rangle_L}{(\rho_2-\rho_{2,c})L} \sim
(t/\tau_B)^{(1+\gamma)/(1+2\gamma)}\;, \quad {\rm i.e.} \quad \beta_2 =
\frac{1+\gamma}{1+2\gamma}\;.
\end{equation}

\subsubsection{Phase C}

In phase C, $\rho_1 >\rho_{1,c}$ and $\rho_2 >\rho_{2,c}$ and both
species condense. While in phases A and B the relationship between the
occupation numbers $k_1$ and $k_2$ was fixed by compatibility with the
bulk dynamics, in phase C this relationship is not uniquely determined.
Using the expansion
\bea\label{ratec}
g_1 (k_1 ,k_2 )\approx 1-b\gamma k_2 /k_1^{1+\gamma} +c/k_1 \ ,\quad
g_2 (k_1 ,k_2 )\approx 1+b/k_1^\gamma \;,
\eea
we see that for $k_1 =\mathcal{O}(L)$ any value of $k_2$ in the range
$\mathcal{O}(1)\leq k_2 
\leq\mathcal{O}(L)$, and for $k_2 =\mathcal{O}(L)$ any value of $k_1$
in the range
$\mathcal{O}(L^{1/(1+\gamma)})<k_1 \leq\mathcal{O}(L)$, lead to $g_1
\approx g_2 \approx 1$ and are compatible with the bulk dynamics.
All compatible relations between $k_1$ and $k_2$ 
may be observed during the coarsening regime, but the sites with the
longest lived relation will determine the coarsening timescale.

Since the leading order of $k_1$ in the hop rate $g_1$ given in
(\ref{ratec}) depends on $\gamma$, we have to distinguish three cases:
\begin{itemize}
\item[$\gamma {<}1$:] In this case the longest lived relation
  is given by double cluster sites, i.e.\ sites with $k_1
  \sim k_2 \sim L$. Since $g_1 -1=-b\gamma k_2 /k_1^{\gamma +1}<0$
  such cluster sites gain excess species 1 particles from the bulk
  rather than losing them.

Double cluster sites are stable compared to cluster sites with other
relationships between $k_1$ and $k_2$, in the sense that such sites are
driven towards $k_1 \sim k_2 \sim L$: For $k_2 =\mathcal{O}(L)$ and
$\mathcal{O}(L^{1/(1+\gamma)})<k_1 <\mathcal{O}(L)$ one has $g_1 (k_1
,L)-g_1 (L,L)\sim -b\gamma L/k_1^{1+\gamma}$. Therefore the smaller
the value of $k_1$ at cluster sites, the greater the rate at which
species 1 particles are gained from the bulk. Thus $k_1$ is driven
towards a value $\mathcal{O}(L)$. On the other hand, for $k_1
=\mathcal{O}(L)$ and $k_2 <\mathcal{O}(L)$ one has $g_1 (L,k_2 )-g_1
(L,L)\sim b\gamma /L^{\gamma} >0$, so cluster sites at which only
$k_1={\mathcal O}(L)$ lose species 1 particles to double cluster
sites.

Since on double cluster sites both species exchange particles with
the bulk at an effective rate proportional to $1/k_i^\gamma$,
$i=1,2$, the coarsening timescale is given by $t(k_i) \sim k_i
k_i^\gamma$. Thus $\tau_C \sim L^{1+\gamma}$ and both species coarsen
with the same exponent $\beta_i=1/(1+\gamma)$, i.e.
\begin{equation} \label{Cscal1}
\frac{\big\langle m_1 (t)\big\rangle_L}{(\rho_1 -\rho_{1,c} )L} \sim
\frac{\big\langle m_2 (t)\big\rangle_L}{(\rho_2
-\rho_{2,c} )L} \sim \big( t/\tau_C \big)^{\frac{1}{1+\gamma}}\; .
\end{equation}
\item[$\gamma {=}1$:] The longest lived sites are again double cluster
  sites, at which $k_1 \sim k_2 \sim L$. Now the sign of $g_1
  -1\approx -b\gamma /L +c/L$ depends on $b$ and $c$, but all the
  arguments for $\gamma <1$ apply in this case also, so we expect that
  the scaling law (\ref{Cscal1}) still holds for $\gamma=1$.

\item[$\gamma {>} 1$:] Now the leading order for cluster sites of the
  first species changes to $g_1 \approx 1+c/k_1$ independent of
  $\gamma$ and $k_2$. Thus species 1 coarsens independently of species
  2 with the dynamics determined in the same way as phase A, therefore
  $\beta_1 = 1/2$. However, the relation between $k_1$ and $k_2$ is not
  stable on cluster sites at which $k_1={\mathcal O}(L)$ for any value
  of $k_2$, since $g_2 -1\approx b/k_1^\gamma <g_1 -1\approx
  c/k_1$. But when $k_1$ is large, $k_2$ is driven towards large
  values (since $g_2 -1$ is small), thus excess species 2 particles
  accumulate at sites where $k_1={\mathcal O}(L)$.
% but the dynamics is on a slower time scale as for the
%  species 1 particles.

%So species 2 particles coarsen after species 1 particles. 
On species 2 cluster sites, i.e.\ sites where $k_2 \sim L$, the slowest
timescale in the dynamics of the species 2 particles is set when the
cluster site contains $k_1 = {\mathcal O}(L)$ species 1 particles.
%However this is not a stable ratio of $k_1$ an $k_2$ which can be seen
%as follows. 
However, since the effective exit rates, $g_1-1$ and $g_2-1$, differ
for each species, the coarsening mechanism is more complicated than in
previous cases. This can be seen as follows.  When $k_1 = {\mathcal
O}(L)$, species 2 particles are lost to the bulk with an effective
rate proportional to $\mathcal{O}(L^{-\gamma})$. Now, the time it
would take for $\mathcal{O}(L)$ species 2 particles to escape to the
bulk scales as $\mathcal{O}(L^{1+\gamma})$ which is large (since
$\gamma>1$) compared to the timescale $\mathcal{O}(L^2)$ over which
the species 1 particles coarsen. Therefore after a time of order
$\mathcal{O}(L^2)$, the number of species 2 particles at a cluster
site is still $\mathcal{O}(L)$ but the number of species 1
particles has decreased to its minimum value allowed by continuity,
$\mathcal{O}(L^{1/(1+\gamma)})$. Now the species 2 particles are lost
to the bulk in a time of order $\mathcal{O}(L^{1+\gamma/(1+\gamma)})$
which is fast relative to the time of order $\mathcal{O}(L^2)$ we have
already waited for the species 1 particles to coarsen.
Thus the species 2 cluster dismantles immediately following the
dissolution of the species 1 cluster. Hence both species coarsen on a
time scale $\tau_C \sim L^2$ and we expect
\begin{equation}
\frac{\big\langle m_1 (t)\big\rangle_L}{(\rho_1 -\rho_{1,c} )L} \sim
\frac{\big\langle m_2 (t)\big\rangle_L}{(\rho_2
-\rho_{2,c} )L} \sim \big( t/\tau_C \big)^{1/2}\; .
\end{equation}
The coarsening of species 2 almost exclusively takes place on
vanishing species 1 cluster sites.  In this sense the coarsening of
the species 2 particles is effectively a slave to that of the species
1 particles. Indeed in simulations this picture is confirmed, and both
species coarsen on the same time scale, but the species 1 particles
coarsen first (see Figure \ref{phasec} in the next section).

\end{itemize}
\begin{table}
\begin{center}
\begin{tabular}{|c|c|}
\hline 
Phase & Coarsening exponents \\ 
\hline \hline 
A & $\beta_1=1/2$\\ 
B & $\beta_2 = \frac{1+\gamma}{1+2\gamma}$ \\ 
C, $\gamma\leq 1$ & $\beta_1 = \beta_2 = \frac{1}{1+\gamma}$ \\
C, $\gamma > 1$ & $\beta_1=\beta_2 = 1/2$ \\
\hline
\end{tabular}
%\hfill\raisebox{-29mm}{\includegraphics[width=0.5\textwidth]{exponents.eps}}
\caption{Coarsening exponents for asymmetric hopping
%(left), plotted as a function of $\gamma$ (right).
} \label{ASYMTAB}
\end{center}
\end{table}
\begin{figure}
\begin{center}
\includegraphics[width=0.5\textwidth]{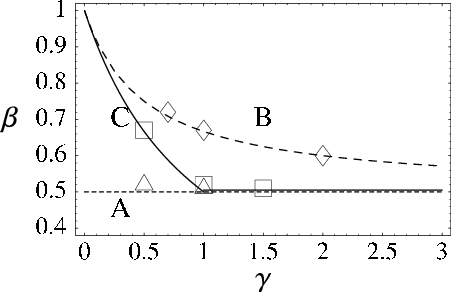}
\caption{Comparison of the theoretical predictions, indicated by
  lines, with the numerical estimates of the exponents obtained in phase A
  ($\triangle$), B ($\diamond$) and C ($\Box$). Errors are of the size
  of the symbols.}
\label{fig:comp}
\end{center}
\end{figure}
The results for each phase are summarized in Table
  \ref{ASYMTAB}. These theoretical predictions are compared with
  numerical results, which are presented in
  the next subsection, in Figure \ref{fig:comp}.

\subsection{Comparison to simulation data}

\begin{figure}
\begin{center}
\includegraphics[width=0.85\textwidth]{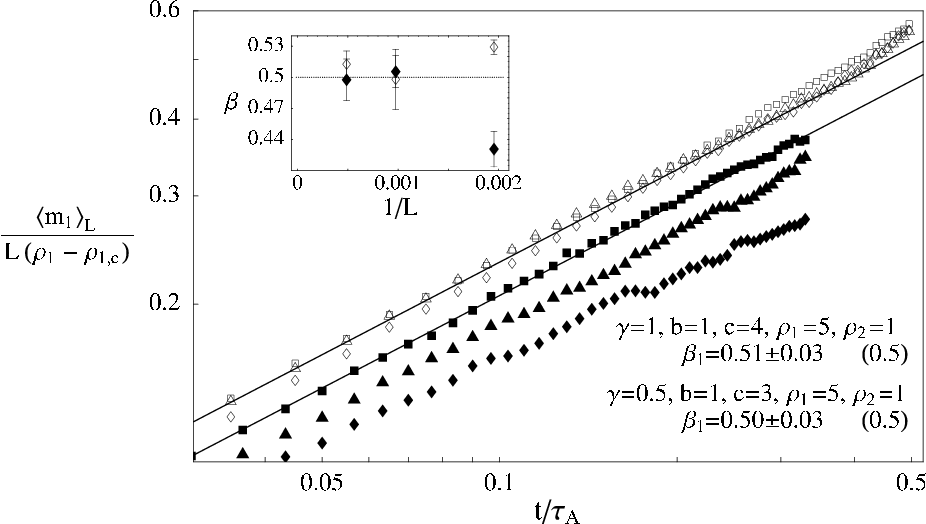}\\[5mm]
\includegraphics[width=0.85\textwidth]{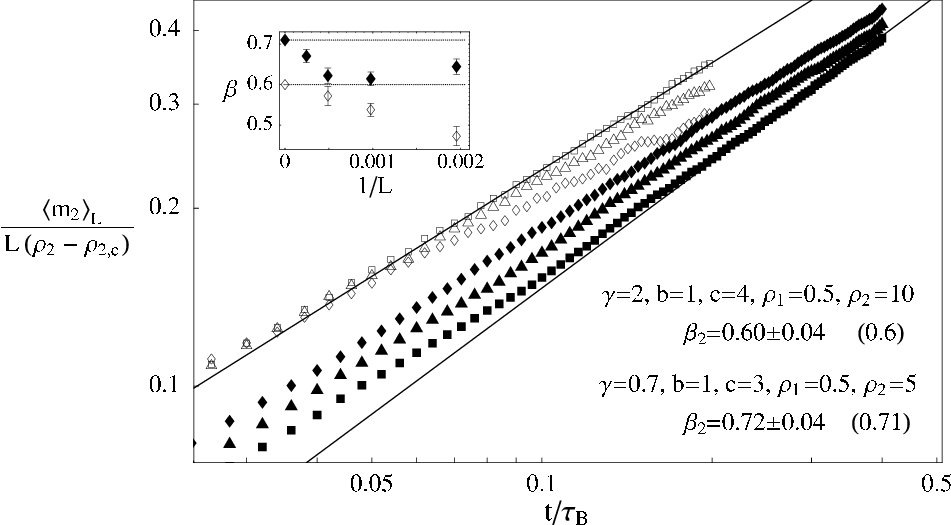}\\
\caption{Verification of the predicted scaling laws, denoted by
  straight lines in a double logarithmic plot of the normalized mean
  condensate size $\tfrac{\langle m_i (t)\rangle_L}{L(\rho -\rho_c )}$
  for asymmetric hopping as a function of time. The predicted scaling
  exponents are given in parentheses. The insets show the finite size
  scaling for the numerical estimates, where filled and unfilled
  symbols correspond to the data.\newline \textbf{Top:} Phase A with
  two sets of parameters. Symbols $L=512 (\Diamond )$, $1024
  (\triangle )$, $2048 (\Box )$ for $\gamma =1$ and filled symbols for
  $\gamma =0.5$.\newline \textbf{Bottom:} Phase B with two sets of
  parameters. Symbols $L=512 (\Diamond )$, $1024 (\triangle )$, $2048
  (\Box )$ for $\gamma =2$ and $L=1024 (\blacklozenge )$, $2048
  (\blacktriangle )$, $4096 (\blacksquare )$ for $\gamma =0.7$}
\label{phaseab}
\end{center}
\end{figure}

\begin{figure}
\begin{center}
\includegraphics[width=0.85\textwidth]{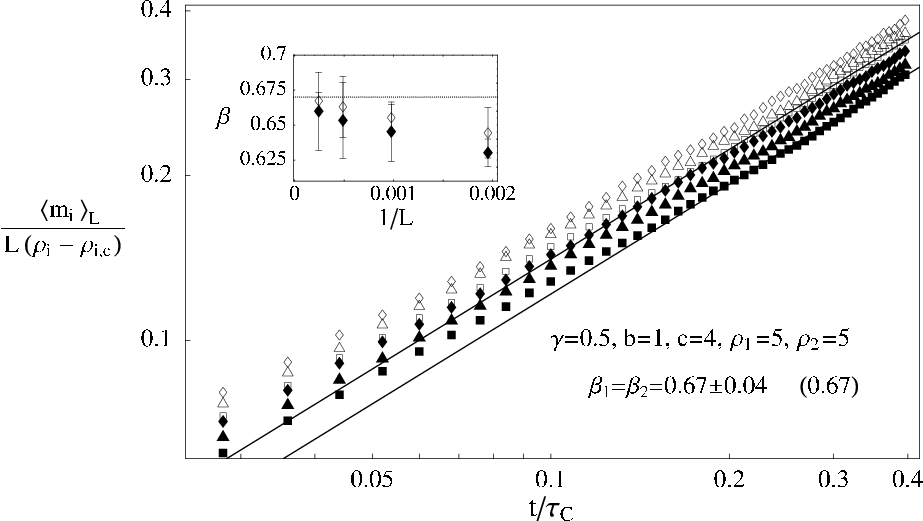}\\[5mm]
\includegraphics[width=0.85\textwidth]{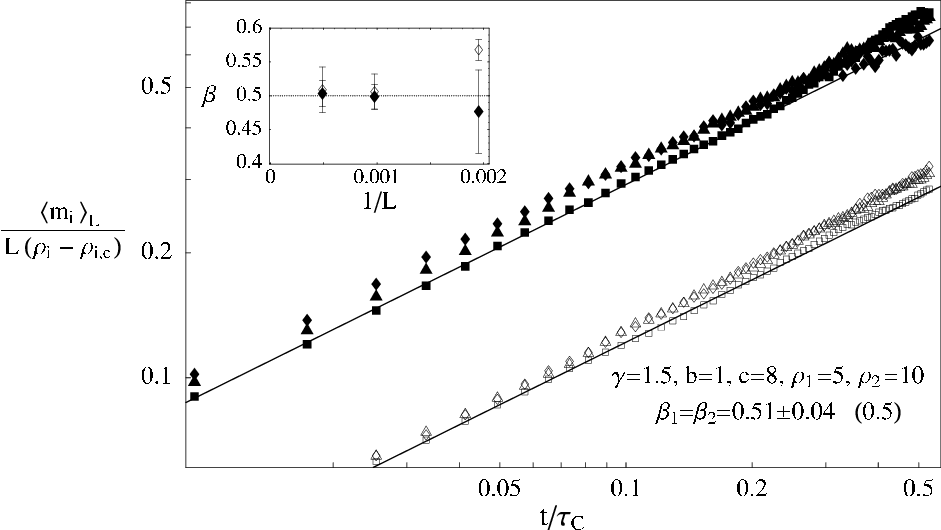}\\
\caption{Verification of the predicted scaling laws in phase C. The
  plots are analogous to Figure \ref{phaseab}, except that each one
  corresponds to only one value of $\gamma$ and condensate sizes of
  both species are shown in filled and unfilled symbols.\newline
\textbf{Top:} Phase C with $\gamma =0.5$. Symbols $L=1024 (\Diamond )$, $2048
  (\triangle )$, $4096 (\Box )$ for species 2 and filled symbols for
  species 1.\newline
\textbf{Bottom:} Phase C with $\gamma =1.5$. Symbols $L=512 (\Diamond
  )$, $1024 (\triangle )$, $2048 (\Box )$ for species 2 and filled symbols for
  species 1.}
\label{phasec}
\end{center}
\end{figure}

The theoretical predictions of the previous subsection are compared to
Monte Carlo simulations in Figures \ref{phaseab} and
\ref{phasec}. $N_1 =[\rho_1 L]$ resp.\ $N_2 =[\rho_2 L]$ particles of
species 1 resp.\ 2 are initially distributed on a lattice of size
$L$ with uniform probability. Cluster sites of species $i$ are defined
by the threshold $(\rho_i -\rho_{i,c})L/40$. The proportionality
factor has to be chosen such that bulk fluctuations are well separated
from cluster sites. Results for exponents are not sensitive to this
choice for the system sizes considered, ranging from $L=256$ to
$4096$, since the fluctuations grow only sublinearly in $L$.  With
this threshold we measure the mean condensate size $m_i (t)$ and other
observables, such as the bulk density $\rho_{i,\mathrm{bulk}}(t)$, of
species $i$ as a function of time, scaled with the expected coarsening
time scale $\tau$. The ensemble average $\langle ..\rangle_L$ is
approximated by averaging over 400 sample runs for each system size.

In the following we discuss the expected behaviour of the observables
which is consistent with simulation results, up to finite size
effects, discussed in the appendix. Plots of the normalized mean
condensate size $\langle m_i (t)\rangle_L /(\rho_i -\rho_{i,c})L$ for
different system sizes $L$ against the rescaled time $t/\tau$, where
$\tau$ is the predicted coarsening time scale, are expected to
collapse onto a single curve.  Within the coarsening regime this curve
should be described by the scaling laws derived in Section IV.

During nucleation, more and more particles become trapped in cluster
sites, therefore the bulk density $\rho_{i,\mathrm{bulk}} (t)$ is a
decreasing function of time, approaching the critical density
$\rho_{i,c}$. This is used as a criterion to identify the beginning of
the coarsening regime. The end of the coarsening regime is reached
approximately when $\langle m_i (t)\rangle_L =0.4\, (\rho_i
-\rho_{i,c})L$, corresponding to an average of $2.5$ remaining cluster
sites. For later times the data significantly deviate from the scaling
law and the system saturates, as already explained in (\ref{sclaw2}).
Within the coarsening time regime defined above we
make a linear fit to the double logarithmic data points of the
normalized mean condensate size, showing an approximately linear
behaviour. The measured slope gives the numerical estimate for the
coarsening exponent $\beta_i$. To get sensible error estimates we
slice the coarsening time window into four smaller time windows (which
may overlap), and measure the exponent in each of the windows. The
error $\beta_i$ is then taken as the standard deviation of these
measurements.


The simulation results are shown in Figure \ref{phaseab} for phases A
and B, and in Figure \ref{phasec} for phase C. We plot
the data on a double logarithmic scale for the three largest system sizes and compare to the
expected scaling law given by straight lines. Finite size scaling of
the measured exponents is given in insets. In phase C the measurements show
rather large errors but are in good agreement with the
predictions. For $\gamma =1.5$ we see that, as expected, the species 1
particles coarsen first, but with the same exponent as species 2
particles. In phases A and B error bars are smaller, but there are
stronger finite size effects affecting the quality of the data collapse.
Nevertheless the measured scaling exponents are in good
agreement with the predictions.

Finite size effects strongly depend on the parameter $\gamma$, and
since there are many competing mechanisms the value or even the sign of the
resulting finite size correction is very hard to estimate. We provide
a discussion of these finite size effects and their influence on the
data collapse in the appendix.

\subsection{Discussion}

\begin{figure}[t]
\begin{center}
\includegraphics[width=0.85\textwidth]{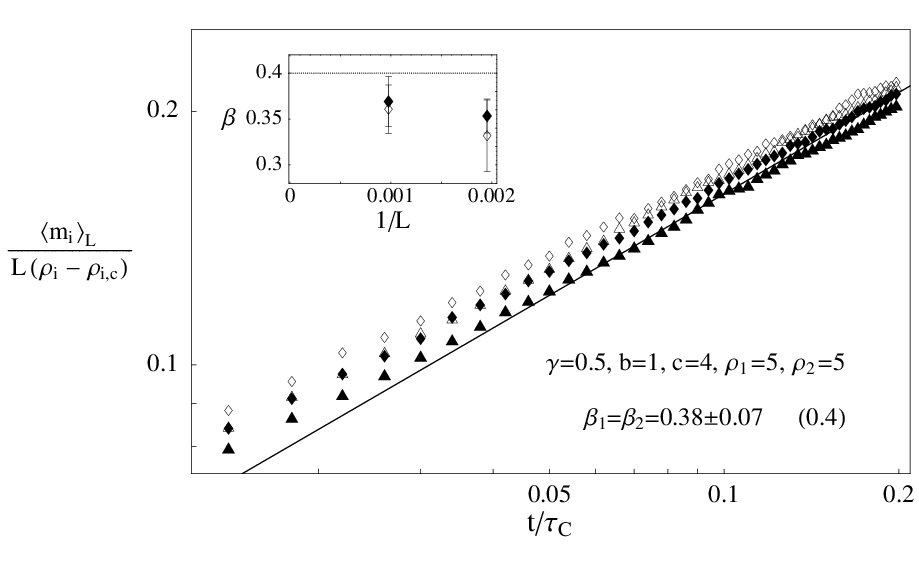}\\
\caption{Verification of the predicted scaling laws for symmetric
  hopping in phase C with $\gamma =0.5$. Details of the plot are given
  in the caption to Figure
  \ref{phaseab}.\newline
Symbols $L=512 (\Diamond )$, $1024 (\triangle )$ for species 2 and
  filled symbols for species 1.}
\label{symmc}
\end{center}
\end{figure}

%obtains the same prediction when the particle hopping is not
%totally but only partially asymmetric, whereas for symmetric hopping
%there is an additional point. 
One can also obtain predictions for the coarsening exponents
when the hopping is symmetric. In this case, there is a high
probability that a particle leaving a cluster site will return to the
same site. The probability that it reaches the next cluster site is
inversely proportional to the distance between cluster sites
\cite{weiss} so it is of order $L^{-1}$. Therefore only every
$\mathcal{O}(L)$-th excess particle will reach the next cluster
site. Hence the coarsening time scales are increased by a factor of
order $L$. The assumption that excess particles move independently
through the bulk remains a good one however: the time it takes
particles to enter the bulk increases by a factor $\mathcal{O}(L)$
(compared with the driven case) since most particles return to the
site they have just left; this cancels the $\mathcal{O}(L)$ increase
in the time particles spend in the bulk due to the diffusive rather
than driven motion. Then the arguments presented for the driven case
with the extra factor $\mathcal{O}(L)$ in the coarsening time scales
leads to the exponents given in Table \ref{SYMTAB}. We compare the
prediction to preliminary simulation data in Figure \ref{symmc}, where
we get good data collapse on the symmetric time scale. The system sizes are,
however, too small for a reasonable numeric estimate of the
scaling exponent, but at least one can see that the exponent is
significantly smaller than for totally asymmetric hopping (cf.\ Figure
\ref{phasec} top).

For partially asymmetric hopping excess particles return to the
cluster site they just left with non-zero probability, but also the
probability of reaching the next cluster site in direction of the
drive is of order 1, even in the limit $L\to \infty$. So in
contrast to symmetric hopping the coarsening time scale is only
corrected by a constant factor independent of $L$, and the coarsening
exponent is the same as in the totally asymmetric case.
\begin{table}
\begin{center}
\begin{tabular}{|c|c|}
\hline 
Phase & Coarsening exponents \\ 
\hline \hline 
A & $\beta_1=1/3$\\ 
B & $\beta_2 = \frac{1+\gamma}{2+3\gamma}$ \\ 
C, $\gamma\leq 1$ & $\beta_1 = \beta_2 = \frac{1}{2+\gamma}$ \\
C, $\gamma > 1$ & $\beta_1=\beta_2 =1/3$ \\
\hline
\end{tabular}
\caption{Coarsening exponents for symmetric hopping.} \label{SYMTAB}
\end{center}
\end{table}

It is interesting to compare our prediction for the exponents in phase
B with the results of \cite{barma} and \cite{jain} in which the
authors study a single-species zero-range process where the hop
rates, $w_1, \ldots, w_L$, are site-dependent but independent of the
particle occupation number at the departure site. They consider
symmetric \cite{barma} and asymmetric \cite{jain} dynamics, where the
(quenched) hop rates are drawn independently from a distribution
$p(w)$ which can be written in the form
\begin{equation} \label{p(w)}
p(w) =
[(\gamma^{-1}+1)/(1-\alpha)^{\gamma^{-1}+1}](w-\alpha)^{\gamma^{-1}}\;,    
\end{equation}
where $w\in [\alpha,1]$ with $\gamma, \alpha>0$. This model undergoes
a condensation transition above a critical particle density from a
homogeneous phase to a phase with a condensate which resides at the
site with the smallest hop rate. In both asymmetric and symmetric
cases, they obtain coarsening exponents identical to those we obtain
for the coarsening of the species 2 particles in phase B. One can
think of the dynamics defined in (\ref{rats}) as a model of particles
(species 2 particles) moving on an evolving disordered background
(given by the species 1 particles). By the time the coarsening regime
has been reached, at the cluster sites the evolving disorder is
effectively quenched. Therefore it is not necessarily surprising that
the two models exhibit similar coarsening behaviour for some
distribution $p(w)$. The reason the form (\ref{p(w)}) is the relevant
one for the rates (\ref{rats}) is as follows. In the disordered model
the coarsening is governed by the exchange of particles between the
two slowest sites in the system. The rate at which particles are
transferred between these two slowest sites is given by the difference
between the two rates at these sites, $\Delta w$. For the distribution
(\ref{p(w)}), $\Delta w \sim L^{-\gamma/(1+\gamma )}$. 
This rate separation contributes the same factor to the coarsening
time scale as that in the two-species model due to the dependence of
the hop rate of species 2 particles on the background of species 1
particles (see equation (\ref{ratesb2})). 
The remaining contributions to the coarsening time
scale are then determined by the symmetry of the hopping dynamics,
i.e.\ the coarsening time scale is given by a factor of order $L$ for
asymmetric dynamics, or a factor of order $L^2$ for symmetric
dynamics, multiplied by the inverse rate separation. This leads to the
same exponents as those obtained for phase B.

\section{Conclusion}

We have considered a two-species zero-range process which undergoes a
variety of transitions to different
condensate phases. The combination of two conservation laws and the
coupling in the dynamics between the two species of particles leads to
coarsening dynamics which are very rich compared to the single species
model. In particular, we have considered a case in which the dynamics
of one of the particle species (the species 2 particles) depends only
on the number of particles of the other species (species 1) at the
departure site,
%. We considered the coarsening dynamics in the case
%where the hop rate for species 2 particles
and decays to a constant value
as a power law with exponent $\gamma$. While the stationary phase
diagram discussed in Section II.B depends also on other system
parameters, the coarsening exponents only depend (continuously) on
$\gamma$ and differ from phase to phase.
Further, as expected, the exponents depend on the symmetry of the
hopping dynamics.
 
{\bf Acknowledgments} The authors thank the Max Planck Institute for
Complex Systems, Dresden, where this work was initiated, for
hospitality. TH was supported by EPSRC programme grant GR/S10377/01.
 
\appendix

\section*{Appendix}
\section*{Discussion of finite size effects}

\begin{figure}[t]
\begin{center}
\includegraphics[width=0.45\textwidth]{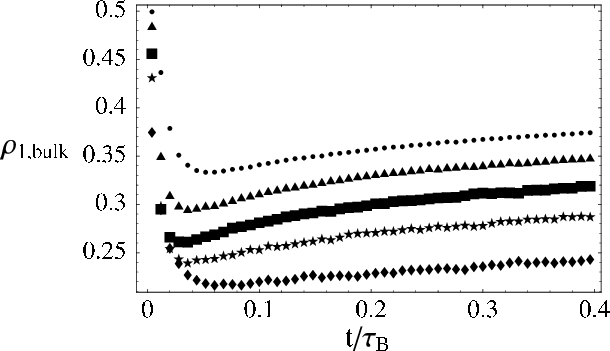}\hfill
\includegraphics[width=0.45\textwidth]{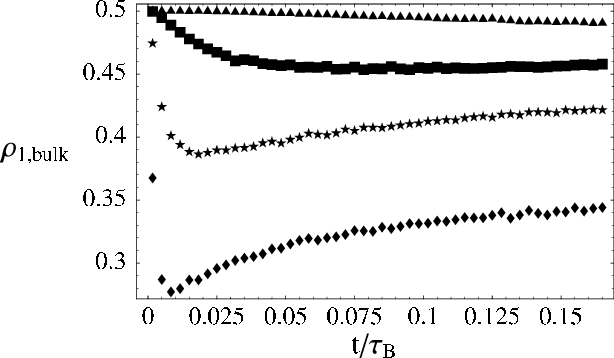}\\[5mm]
\includegraphics[width=0.45\textwidth]{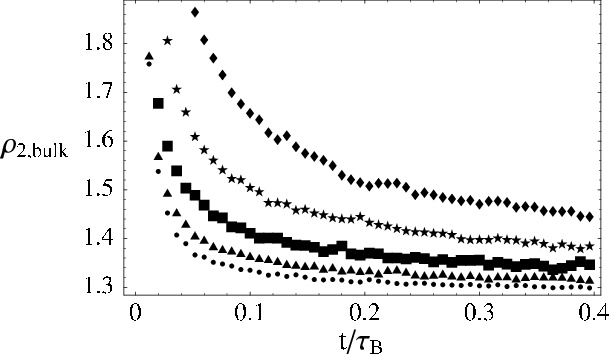}\hfill
\includegraphics[width=0.45\textwidth]{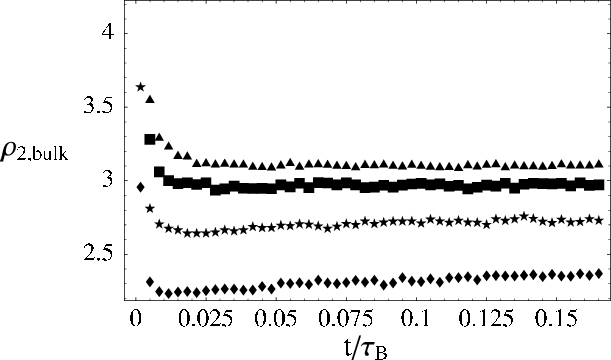}\\
\hspace*{10mm}
$\gamma =0.7$, $\rho_1 =0.5$,
$\rho_{2,c}=1.28$\hspace*{42mm} $\gamma =2$, $\rho_1 =0.5$,
$\rho_{2,c}=3.40$\\ 
\caption{Finite size effects in phase B. $\rho_{1,\mathrm{bulk}}$ shown
on the top is smaller than its limiting value $\rho_1 =0.5$ for
$L\to\infty$ since on species 2 cluster sites $k_1 \sim L^{1/(1+\gamma
)}$. With increasing $\gamma$ this effect becomes
weaker. $\rho_{2,\mathrm{bulk}}$ shown on the bottom converges to its
limiting value $\rho_{2,c}$ from above (for $\gamma <1$) and from
below (for $\gamma >1$) as explained in the text.\newline
\textbf{Left:} $\gamma =0.7$, $b=1$, $c=3$, $\rho_1 =0.5$, $\rho_2
=5$, critical density $\rho_{2,c}=1.28$\newline Symbols: $L=256
(\blacklozenge )$, $512 (\bigstar )$, $1024 (\blacksquare )$, $2048
(\blacktriangle )$, $4096 (\bullet )$\newline \textbf{Right:} $\gamma
=2$, $b=1$, $c=4$, $\rho_1 =0.5$, $\rho_2 =10$, critical density
$\rho_{2,c}=3.40$\newline Symbols: $L=256 (\blacklozenge )$, $512
(\bigstar )$, $1024 (\blacksquare )$, $2048 (\blacktriangle )$}
\label{finitesize}
\end{center}
\end{figure}

In the following we discuss qualitatively the most basic mechanisms
leading to finite size effects and illustrate how they depend on the
system parameters. We start with two competing effects which influence
the bulk density of the condensing species, connected with the
definition of cluster sites by a threshold. (i) Excess particles
exchanged between cluster sites increase the bulk density in finite
systems. In phase B, for example, this effect is of order
$L^{1-\gamma/(1+\gamma )} =L^{1/(1+\gamma )}$, by reasoning given in Section
III.B and the expansion (\ref{ratesb2}). It decreases with
increasing $\gamma$ and leads to a decrease of the mean condensate
size. This effect is shown for $\gamma <1$ 
in Figure \ref{finitesize} (bottom left), where
$\rho_{2,\mathrm{bulk}}$ is plotted, decreasing towards $\rho_{2,c}$
with increasing system size $L$. (ii) On the other hand, bulk
fluctuations increase with $\gamma$ and in finite systems they can
exceed the threshold for cluster sites. This leads to an increase of
the number of condensed particles, or a decrease of
$\rho_{2,\mathrm{bulk}}$, which dominates over effect (i) for $\gamma
>1$. This can be seen in Figure \ref{finitesize} (bottom right) where
$\rho_{2,\mathrm{bulk}}$ now increases towards $\rho_{2,c}$. Since
these fluctuations are typically small compared to other cluster
sites, this leads also to a decrease of the mean condensate
size.
(iii) Further, in finite systems the nucleation and the
coarsening regime are not clearly separated but overlap to a large
extent, as can also be seen in Figure \ref{finitesize}
(bottom). Ongoing nucleation effectively slows down the coarsening,
leading to a decrease of the coarsening exponents for finite system
sizes, as can be seen in the insets of Figures 3 (bottom) and 4 (top).
% and we expect an exponential correction term for the scaling law,
%\bea
%\frac{\langle m_i (t)\rangle_L}{(\rho_i -\rho_{i,c})L}\sim
%\Big(\frac{t}{\tau}\Big)^{\beta_i} \Big( 1+C_1 e^{-\frac{\lambda}{L}t}\Big)
%=\Big(\frac{t}{\tau}\Big)^{\beta_i} \Big( 1+C_1 e^{-C_2 L^\alpha \frac{t}{\tau}}\Big)\ ,
%\eea
%where nucleation takes place on a time scale of order $L$ and the coarsening
%time scale is $\tau\sim L^{1+\alpha}$.
%This affects the data collapse and leads to a decrease of the measured
%coarsening exponents for finite system sizes. The effect is larger for
%small values of $\gamma$. In Figure \ref{regioncc_corrected} we show a plot in phase C, $\gamma
%=0.5$ of the data corrected by the factor $\displaystyle\Big( 1+C_1 e^{-C_2
%  L^\alpha \frac{t}{\tau}}\Big)^{-1}$ with fitting parameters $C_1$ and
%$C_2$. We see a drastic improvement in the data collapse and in the finite size
%corrections for the exponents. The same can be done for $\gamma =1.5$ where
%the corrections are much smaller, whereas in phase A and B other effects are
%sometimes dominating (cf.\ Figure \ref{phaseab}).
\begin{figure}[t]
\begin{center}
\includegraphics[width=0.85\textwidth]{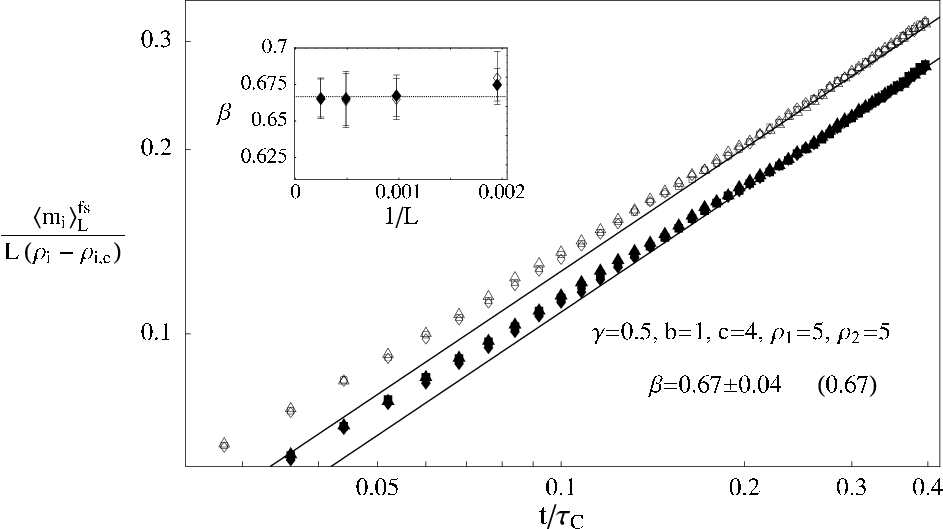}\\
\caption{Finite size corrected data for phase C with $\gamma =0.5$ as
  given in (\ref{corrdata}).}
\label{regioncc_corrected}
\end{center}
\end{figure}

On top of these corrections, which are also present in single species
models, there are mechanisms specific to two-component systems. (iv)
The coarsening dynamics does not only depend on the occupation number
of the condensing species, but on the relation between $k_1$ and
$k_2$. As discussed above, condensates with different relations have
different lifetime. So in the limit $L\to\infty$ ratios differing by
some factor $L^\alpha$ are dynamically separated and only one relation
dominates the coarsening. But for finite systems, ratios with shorter
lifetimes also contribute to the observed behaviour. Thus data for
single species systems, where this effect does not occur, are
generically better than our data (cf.~\cite{GSS}).  (v) Finally, we
consider a phenomenon specific for phase B. As discussed above, on
species 2 cluster sites there are of order $L^{1/(1+\gamma )}$ species
1 particles due to compatibility with the bulk. For $\gamma <1$ this
is larger than a typical bulk fluctuation of order $L^{1/2}$, and
reduces the bulk density of species 1 particles, whereas for $\gamma
>1$ the effect is much weaker. This can be seen in Figure
\ref{finitesize} (top), where $\rho_{1,\mathrm{bulk}}$ is plotted.

The variety of effects leads to a diverse behaviour and a quantitative
prediction of the finite size corrections seems to be not
feasible. However, it is straightforward to numerically fit the
leading order corrections for the prefactor and the exponent of the
scaling law (\ref{sclaw}), using the ansatz
%in the slope and the shift in the double
%logarithmic data,
%two errors,
%a change of the exponent which is proportional to $1/L$ and a shift of the data.
\bea
\frac{\big\langle m_i(t) \big\rangle_L }{(\rho_i -\rho_{i,c})L}=C_1\big( 1
+C_2 /L^{\delta_1}\big)\Big(\frac t\tau\Big)^{\beta_i +C_3
  /L^{\delta_2}}\ .
\eea
Consider for instance the data in phase C with $\gamma =0.5$ given in
Figure \ref{phasec} (top). The finite size corrections in the inset
suggest $\delta_2 =1$ and the best fit values for the other parameters are
$\delta_1 =0.69,\ C_1 =0.51,\ C_2 =2.3,\ C_3 =-21$ for species 1 and
$\delta_1 =0.60, C_1 =0.59,\ C_2 =2.1,\ C_3 =-13$ for species 2. In
Figure \ref{regioncc_corrected} we plot the corrected data
\bea\label{corrdata}
\frac{\big\langle m_i(t) \big\rangle_L^{fs} }{(\rho_i -\rho_{i,c})L}
=\frac{\big\langle m_i(t) \big\rangle_L }{(\rho_i -\rho_{i,c})L}
\Big(\frac t\tau\Big)^{-C_3 /L^{\delta_2}} \big( C_1 +C_2
/L^{\delta_1}\big)^{-1}\ ,
\eea
and the data collapse improves drastically.

\end{document}